\documentstyle[aps,prl,epsfig,float,twocolumn]{revtex}
\begin{document}
\twocolumn[\hsize\textwidth\columnwidth\hsize\csname
@twocolumnfalse\endcsname

\title{Entanglement in SU(2)-invariant quantum spin systems}

\author{John Schliemann}

\address{Department of Physics and Astronomy, University of Basel,
CH-4056 Basel, Switzerland}

\date{\today}

\maketitle

\begin{abstract}
We analyze the entanglement of SU(2)-invariant density matrices of two
spins $\vec S_{1}$, $\vec S_{2}$ using the Peres-Horodecki criterion. 
Such density matrices arise from thermal equilibrium states of 
isotropic spin systems. The partial transpose 
of such a state has the same multiplet structure and degeneracies 
as the original matrix with the eigenvalue of largest multiplicity being
non-negative.
The case $S_{1}=S$, $S_{2}=1/2$ can be solved completely and is discussed 
in detail with respect to isotropic 
Heisenberg spin models. Moreover, in this case the Peres-Horodecki 
ciriterion turns out to be a sufficient condition for non-separability.
We also characterize SU(2)-invariant states of
two spins of length 1.
\end{abstract}
\vskip2pc]

\section{Introduction}

Entanglement is one of the most intriguing properties of quantum physics
\cite{Einstein35,Schrodinger35}
and the key ingredient of the emerging field of quantum information
theory and processing \cite{Nielsen00}. Very recently substantial interest
has developed concerning the question of quantum entanglement in
equilibrium states of quantum spin systems 
\cite{Wootters00,OConnor01,Arnesen01,Gunlycke01,Wang01,Wang02a,Wang02b,Wang02c,Kamta02,Osterloh02,Osborne02,Bose02,Meyer02,Vidal02,Latorre03,Jin03}
as often studied in condensed matter physics and statistical mechanics. 
More specifically,
a typical question arises: Tracing out from a many-body system all degrees of
freedom except for, say, two spins, is this reduced density matrix
separable or not?

In the present work we study SU(2)-invariant density matrices
of two spins. These states are are defined to be 
invariant under all uniform rotations $U_{1}\otimes U_{2}$ of both spins
$\vec S_{1}$ and $\vec S_{2}$, where $U_{a}=\exp(i\vec\eta\vec S_{a})$,
$a\in\{1,2\}$, are transformations
corresponding to the same set of real parameters $\vec\eta$
in the representation of SU(2) appropriate for the spin lengths 
$S_{1}$ and $S_{2}$ ($\hbar=1$). In other words, such states 
$\rho$ commute with all components of the total spin 
$\vec J=\vec S_{1}+\vec S_{2}$.
SU(2)-invariant density matrices arise from thermal equilibrium states of
spin systems with an rotationally invariant Hamiltonian 
by tracing out all degrees of freedom but those two spins \cite{note1}.
Our results generalize previous work on such systems
\cite{Wootters00,OConnor01,Wang02b,Wang02c} to the case of higher spins,
and we discuss our findings with respect to generic 
antiferromagnetic or ferrimagnetic Heisenberg spin lattice models.
These considerations lead to the natural conclusion that pairwise quantum
entanglement in equilibrium states is strongest in systems with
small spin length and low spatial dimension. Moreover, an inseparable
equilibrium reduced two-spin state can usually only be achieved
for neighboring spins, but not for more distant lattice sites.  
Viewed like this, equilibrium states of such systems
do not appear to be a particularly strong source of pairwise 
quantum entanglement.   

To investigate the separability of this type of density matrices in the
case of higher spins we shall make use
of the Peres-Horodecki criterion \cite{Peres96,Horodecki96}. This
criterion states that a separable density matrix has necessarily
a positive partial transpose \cite{Peres96}. Moreover a positive
partial transpose is also sufficient \cite{Horodecki96}
for the separability of
a given density matrix in the case of two qubits, and in the case
of a qubit and a qutrit (i.e. the case of two spins of length 1/2, and
of a spin 1/2 and a spin 1, respectively). For larger dimensions
of the parties (spins) involved non-separable states with positive
partial transpose exist \cite{Horodecki97}. To apply the Peres-Horodecki
criterion one can perform the partial transposition with respect to 
either subsystem (spin), since both resulting matrices have the same
spectrum. For definiteness we will consider in the following the 
partial transpose $\rho^{T_{2}}$ with respect to $\vec S_{2}$.

SU(2)-invariant two-spin states where mentioned briefly
already by Vollbrecht and Werner 
\cite{Vollbrecht01}, where it was pointed out that the
case $S_{1}=S_{2}=1$ corresponds to states invariant under 
$O\otimes O$ with $O$ being a O(3) rotation. In this work 
we will also give explicit criteria in terms of spin correlators
for such states to have a positive partial transpose  \cite{noteadded}.

\section{SU(2)-invariant states and their partial transpose}

Let us start with some general considerations. Since an SU(2)-invariant
state commutes with all components of $\vec J$, it acts, according to 
Schur's Lemma, as a scalar on each irreducible representation (multiplet)
of $\vec J$. Therefore $\rho$ has the general from
\begin{equation}
\rho=\sum_{J=|S_{1}-S_{2}|}^{S_{1}+S_{2}}
\frac{A(J)}{2J+1}\sum_{J^{z}=-J}^{J}
|J,J^{z}\rangle_{0}{_{0}\langle J,J^{z}|}\,,
\end{equation} 
where the constants $A(J)$ fulfill $A(J)\geq 0$, $\sum_{J}A(J)=1$.
Here $|J,J^{z}\rangle_{0}$ denotes a state 
of total spin $J$ and $z$-component $J^{z}$.

Now let ${\cal O}$ be a general operator acting on a bipartite system. 
If ${\cal O}$ is transformed by $U_{1}\otimes U_{2}$, ${\cal O}^{T_{2}}$
transforms covariantly under $U_{1}\otimes U_{2}^{*}$, 
\begin{eqnarray}
 & & \left(\left(U_{1}\otimes U_{2}\right){\cal O}
\left(U_{1}^{+}\otimes U_{2}^{+}\right)\right)^{T_{2}}\nonumber\\
 & & =\left(U_{1}\otimes U_{2}^{*}\right){\cal O}^{T_{2}}
\left(U_{1}^{+}\otimes (U_{2}^{*})^{+}\right)\,.
\label{transrel}
\end{eqnarray}
Here $U_{1}$, $U_{2}$ are general unitary transformations acting on the
subsystems and do not necessarily represent SU(2) transformations.
The relation (\ref{transrel}) can be derived readily by writing ${\cal O}$
in the form ${\cal O}_{i_{1}i_{2},j_{1}j_{2}}$ where the subscripts at the
indices refer to the different subsystems \cite{Vollbrecht01,note2}.
In particular, it follows
that if ${\cal O}$ is invariant under $U_{1}\otimes U_{2}$,
${\cal O}^{T_{2}}$ is invariant under $U_{1}\otimes U_{2}^{*}$.

Now let $U_{1}$, $U_{2}$ represent again SU(2) transformations. 
In the standard representations of their generators the $x$- and $z$-components
are given by real matrices while the matrices for the $y$-components
are imaginary. Thus, a complex conjugation of 
$U_{2}=\exp(i\vec\eta\vec S_{2})$,
is equivalent to changing the sign of $S^{x}_{2}$, $S^{z}_{2}$.
Therefore, $\rho^{T_{2}}$ commutes with the operators $\vec K$ defined
by $K^{x}=S^{x}_{1}-S^{x}_{2}$, $K^{y}=S^{y}_{1}+S^{y}_{2}$,
$K^{z}=S^{z}_{1}-S^{z}_{2}$, and these operators also furnish a
representation of su(2), $[K^{\alpha},K^{\beta}]=
i\varepsilon^{\alpha\beta\gamma}K^{\gamma}$ (using standard notation).
In the basis of tensor product states $|S^{z}_{1},S^{z}_{2}\rangle$
of $S^{z}_{1}$, $S^{z}_{2}$, $\rho^{T_{2}}$ is block-diagonal
with respect given values of $K^{z}$. In particular, 
$|\pm S_{1},\mp S_{2}\rangle$ are eigenstates of $\rho^{T_{2}}$ with
the degenerate eigenvalue 
$\langle\pm S_{1},\mp S_{2}|\rho|\pm S_{1},\mp S_{2}\rangle\geq 0$. 
Now it follows from elementary representation theory that $\rho^{T_{2}}$
has actually an SU(2) multiplet structure with respect to the
operators $\vec K$. The multiplets are labeled by the value of
$\vec K^{2}=K(K+1)$ with $|S_{1}-S_{2}|\leq K\leq(S_{1}+S_{2})$.
On these multiplets $\rho^{T_{2}}$ acts as a constant.
As seen above, the eigenvalue corresponding to the largest $K$-multiplet
is always nonnegative.

\section{The case $S_{2}=1/2$}

Let us now consider system consisting of a spin
$\vec S_{1}$ of arbitrary length $S$ and a spin $\vec S_{2}$ of length 
$1/2$. Here a general SU(2)-invariant density matrix has the form
\begin{eqnarray}
\rho & = & \frac{F}{2S}\sum_{J^{z}=-S+1/2}^{S-1/2}
|S-\frac{1}{2},J^{z}\rangle_{0}{_{0}\langle S-\frac{1}{2},J^{z}|}\nonumber\\
 & + & \frac{1-F}{2S+2}\sum_{J^{z}=-S-1/2}^{S+1/2}
|S+\frac{1}{2},J^{z}\rangle_{0}{_{0}\langle S+\frac{1}{2},J^{z}|}\,.
\end{eqnarray}
The quantity $F\in[0,1]$ is in thermal equilibrium 
a function of temperature and, in the case of $\rho$ being a reduced density
matrix of a larger system, it
contains information about the entire system which has been traced out
except for the spins $\vec S_{1}$ and $\vec S_{2}$.
By expressing $F$ in terms of the projector onto the $J=S-1/2$
multiplet one finds
$F=(S-2\langle\vec S_{1}\vec S_{2}\rangle)/(2S+1)$,
where $\langle\cdot\rangle$ denotes the expectation value with respect 
to $\rho$. Thus, $\rho$ is completely determined by the correlator 
$\langle\vec S_{1}\vec S_{2}\rangle$.

In order to perform a partial transposition on $\rho$ it is
convenient to express it in a basis of tensor product eigenstates
$|S^{z},\pm 1/2\rangle$
of $S_{1}^{z}$ and $S_{2}^{z}$. Using the well-known Clebsch-Gordan
coefficients for coupling a spin $S$ to a spin $1/2$, the nonvanishing
matrix elements are given by
\begin{eqnarray}
 & & \langle S^{z},\pm1/2|\rho|S^{z},\pm1/2\rangle\nonumber\\ 
 & & =\frac{1}{2S+1}\left(\frac{(S\mp S^{z})F}{2S}
+\frac{(S\pm S^{z}+1)(1-F)}{2S+2}\right)\\
& & \langle S^{z},1/2|\rho|S^{z}+1,-1/2\rangle\nonumber\\ 
 & & =\frac{\sqrt{(S-S^{z})(S+S^{z}+1)}}{2S+1}
\left(-\frac{F}{2S}
+\frac{1-F}{2S+2}\right)\,.
\end{eqnarray}
The partial transpose $\rho^{T_{2}}$ is diagonal on the subspace
spanned by $|S,-1/2\rangle$, $|-S,+1/2\rangle$ lying in the largest 
$\vec K$ multiplet with the eigenvalue  
$\lambda_{1}:=
\langle S,-1/2|\rho|S,-1/2\rangle=\langle -S,1/2|\rho|-S,1/2\rangle$
where
\begin{equation}
\lambda_{1}=\frac{1}{2S+1}\left(F+\frac{1-F}{2S+2}\right)\,.
\end{equation}
On the remaining Hilbert
space the partial transpose is block-diagonal, where the blocks act on
subspaces spanned by the basis vectors 
$|S^{z},-1/2\rangle$, $|S^{z}+1,1/2\rangle$ and have the from
\begin{eqnarray*}
\left(
\begin{array}{cc}
\langle S^{z},-1/2|\rho|S^{z},-1/2\rangle &
\langle S^{z},1/2|\rho|S^{z}+1,-1/2\rangle\\
\langle S^{z},1/2|\rho|S^{z}+1,-1/2\rangle &
\langle S^{z}+1,1/2|\rho|S^{z}+1,1/2\rangle
\end{array}
\right)\,.
\label{matrix1}
\end{eqnarray*}
The eigenvalues of these submatrices are given by $\lambda_{1}$ and
\begin{equation}
\lambda_{2}=\frac{1}{2S+1}-\frac{1}{2S}F\,.
\end{equation}
These eigenvalues do not depend on $S^{z}$. Therefore
$\lambda_{1}$ and $\lambda_{2}$ occur with the multiplicities $2S+2$ 
and $2S$, respectively, in accordance with the above general results.
Moreover, $\lambda_{1}$ is always positive, while
$\lambda_{2}$ becomes negative for $F>2S/(2S+1)$, or, equivalently,
\begin{equation}
\langle\vec S_{1}\vec S_{2}\rangle<-\frac{S}{2}\,.
\label{crit}
\end{equation}
Thus, our state has non-positive partial transpose if and only if
the correlator $\langle\vec S_{1}\vec S_{2}\rangle$ is negative and
larger in modulus than $S/2$. This is the maximum value 
$|\langle\vec S_{1}\vec S_{2}\rangle|$ 
achieve in a separable state. This intuitively very reasonable
criterion includes earlier results by Wang and Zanardi \cite{Wang02c}
who investigated the case $S_{1}=S_{2}=1/2$ by evaluating the entanglement
of formation \cite{Bennett96} using Wootters' concurrence \cite{Wootters98}.
Unfortunately this is not a viable route for $S_{1}>1/2$ since Wootters'
construction appears to be restricted to the case of two qubits.
Moreover, with increasing $S$, the states with non-positive
partial transpose have increasing weight in the smaller multiplet
$J=S-1/2$, approaching unity for $S\to\infty$.

If a given state has a negative partial transpose it is necessarily entangled.
Moreover, in the case of SU(2)-invariant states with $S_{2}=1/2$
studied in this section, a positive partial transpose, i.e.
$\langle\vec S_{1}\vec S_{2}\rangle\geq-S/2$, is also a sufficient
criterion for separable states. We prove this fact by explicitly
constructing a decomposition consisting of projectors on
porduct states. If a given state has a positive partial transpose we can write
$\langle\vec S_{1}\vec S_{2}\rangle=(S/2)\cos(\gamma)$ with some real angle
$\gamma$. Now let $|0\rangle$ denote a spin-coherent state 
\cite{Radcliffe71,Auerbach94} of $S_{1}$ 
pointing in some arbitrary direction, and $|\gamma\rangle$ denote a 
spin-coherent
state of $S_{2}$ with its polarization direction forming the angle
$\gamma$ with the polarization direction of $S_{1}$. In the pure
product state $|0\rangle\otimes|\gamma\rangle$ we have by construction
$\langle\vec S_{1}\vec S_{2}\rangle=(S/2)\cos(\gamma)$, and the value of
this correlator is invariant under all uniform
rotations of both spins. Now consider
\begin{eqnarray}
\rho & = & N(S)\int d^{3}\eta\Big[
\left(U_{1}(\eta)\otimes U_{2}(\eta)\right)
\nonumber\\
& & \left(|0\rangle\otimes|\gamma\rangle
\langle 0|\otimes\langle\gamma|\right)
\left(U_{1}(\eta)\otimes U_{2}(\eta)\right)^{+}\Big]
\label{decomp}
\end{eqnarray}
where the integration goes over all simultaneous
rotations  parametrized as $U_{a}=\exp(i\vec\eta\vec S_{a})$,
$a\in\{1,2\}$, and $N(S)$ is a normalization constant. 
Thus the state (\ref{decomp}) is a separable state which is obviously 
invariant under simultaneous rotations of both spins and fulfills
$\langle\vec S_{1}\vec S_{2}\rangle=(S/2)\cos(\gamma)$. Since such an 
SU(2)-invariant state is uniquely determined by this correlator, we have 
constructed a decomposition of the original state in terms of
projectors on product states, which completes the proof.

Let us now discuss the above result with respect to isotropic
Heisenberg lattice spin models as 
studied intensively in condensed matter physics and 
statistical mechanics. Previous studies have concentrated on
one-dimensional systems. This has on the one hand the practical reason that
for such systems the body of exact results concerning correlations
is largest. On the other hand this is due to the fact that quantum 
correlations can generically be expected to become weaker with increasing
spatial dimension, i.e. with increasing number of neighbors to each spin.
Therefore one-dimensional systems are the most attractive to look
for equilibrium quantum entanglement.

Since quantum correlations like 
$\langle\vec S_{1}\vec S_{2}\rangle$ can generally be expected to
decay with increasing temperature, the criterion (\ref{crit}) defines
implicitly a threshold temperature for the occurrence of a non-positive
partial transpose, provided the inequality (\ref{crit}) is fulfilled
in the ground state at $T=0$ \cite{Wang02c}. This can only be the case
in antiferromagnetic or, for $S_{1}>S_{2}=1/2$, ferrimagnetic systems.
In particular, in an antiferromagnetic spin 1/2 chain correlations
are generically of the from
\begin{equation}
\langle\vec S_{m}\vec S_{m+n}\rangle=(-1)^{n}\chi(n)\,,
\end{equation}
where $m$ denotes some lattice site in the translationally invariant chain,
$n$ is the number of lattice sites between the spins considered,
and $\chi$ is a positive and monotonously decaying function.
The alternating sign resembles N\'eel ordering as it is present in 
the ground state of a classical antiferromagnet. It follows that
the reduced equilibrium two-spin density matrix can only be entangled
if the spins involved reside on different sublattices (corresponding to
odd $n$). This intuitively clear finding also holds for generic 
antiferromagnets or ferrimagnets on bipartite lattices in higher spatial 
dimension.

The spin 1/2 Heisenberg chain with antiferromagnetic exchange
between nearest neighbors is described by the Hamiltonian
${\cal H}=\sum_{m}\vec S_{m}\vec S_{m+1}$
Here the correlator $\langle\vec S_{m}\vec S_{m+1}\rangle$.
is equal to the ground state energy per spin and given by
$\ln 2-1/4\approx -0.443<-0.25$ \cite{Bethe31}. 
Thus, the criterion (\ref{crit})
is fulfilled \cite{Wang02c}. However, for larger distances between the spins,
$n\in\{3,5,\dots\}$, numerical data \cite{Betsuyaku86} shows that
the inequality (\ref{crit}) is violated and,
according to the Peres-Horodecki criterion, the corresponding reduced
density matrix is separable. The same statements apply to another
typical antiferromagnetic spin 1/2 wave function, the so-called Gutzwiller
wave function \cite{Gutzwiller63}, 
where $\langle\vec S_{m}\vec S_{m+n}\rangle$ can be 
evaluated analytically for all $n$ \cite{Metzner87}.

In summary, one-dimensional anti\-fer\-ro\-mag\-ne\-tic 
iso\-tro\-pic Heisenberg models
of spins 1/2
do generically not appear to be in thermal equilibrium a particularly strong 
source of entanglement, since usually only the reduced density matrices
of neigboring spins are inseparable, while all others are 
non-entangled. This result might appear somewhat surprising
since such systems are usually considered to have particularly
strong spin correlations because of the small spin length as well
as the low spatial dimension \cite{Auerbach94}. However, as seen here,
these strong quantum correlations do in general not translate to
long-ranged entanglement in equilibrium reduced density matrices.

As pointed out already, this finding cannot be expected to change
in the case of higher spatial dimension. Moreover, ferrimagnetic systems
involving spins $S_{1}>1/2$ will generically have the same properties,
since we have the same type of criterion (\ref{crit})
for the partial transpose being non-positive. This criterion 
requires sufficiently strong quantum fluctuations which are generically
reduced with increasing spin length. To illustrate these trends
let us rewrite the criterion (\ref{crit}) in the form
$\langle\vec S_{1}\vec S_{2}\rangle/(S_{1}S_{2})<-1$. For a ferrimagnetic chain
consisting of alternating spins $S_{1}=1$, $S_{2}=1/2$ a numerical
estimate for the l.h.s. of this inequality in the case of neighboring spins
is given by \cite{Brehmer97}
$\langle\vec S_{1}\vec S_{2}\rangle/(S_{1}S_{2})=-1.455$. 
For a two-dimensional spin-1/2 antiferromagnet on the
square lattice one finds \cite{Barnes88}
$\langle\vec S_{1}\vec S_{2}\rangle/(S_{1}S_{2})=-1.344$. Both values
are larger than the result for the spin-1/2 chain as discussed above,
$\langle\vec S_{1}\vec S_{2}\rangle/(S_{1}S_{2})=-1.773$, indicating
the suppression of pairwise entanglement with increasing spatial dimension
and lengths of spins involved.

\section{The case $S_{2}\geq 1$}

We now turn to the case $S_{1}=:S\geq 1$, $S_{2}=1$. Here the general
SU(2)-invariant density matrix reads
\begin{eqnarray}
\rho & = & \frac{G}{2S-1}\sum_{J^{z}=-S+1}^{S-1}
|S-1,J^{z}\rangle_{0}{_{0}\langle S-1,J^{z}|}\nonumber\\
& + & \frac{H}{2S+1}\sum_{J^{z}=-S}^{S}
|S,J^{z}\rangle_{0}{_{0}\langle S,J^{z}|}\nonumber\\
& + & \frac{1-G-H}{2S+3}\sum_{J^{z}=-S-1}^{S+1}
|S+1,J^{z}\rangle_{0}{_{0}\langle S+1,J^{z}|}\,.
\end{eqnarray}
Expressing the quantities $G$ and $H$ in terms of projectors onto the
multiplets on the total spin $J\in\{S-1,S,S+1\}$ one finds
\begin{eqnarray}
G & = & \frac{1}{S(2S+1)}\left(-S-(S-1)\langle\vec S_{1}\vec S_{2}\rangle
+\langle(\vec S_{1}\vec S_{2})^{2}\rangle\right)\\
H & = & 1 -\frac{1}{S(S+1)}\left(\langle\vec S_{1}\vec S_{2}\rangle
+\langle(\vec S_{1}\vec S_{2})^{2}\rangle\right)\,.
\end{eqnarray}
Therefore, $\rho$ is completely determined by the correlators
$\langle\vec S_{1}\vec S_{2}\rangle$ 
and $\langle(\vec S_{1}\vec S_{2})^{2}\rangle$ and has three different
eigenvalues with degeneracies $2S-1$, $2S+1$, and $2S+3$, corresponding to
multiplets of the total spin $\vec J$.
The three different eigenvalues of $\rho^{T_{2}}$ can be found
in the subspaces spanned by 
$(|S^{z}+1,1\rangle,|S^{z},0\rangle,|S^{z}-1-1\rangle)$,
$|S^{z}|\neq S$ having $K^{z}=S^{z}$. With respect to this
basis $\rho^{T_{2}}$ reads
\begin{equation}
\left(
\begin{array}{ccc}
\alpha(S^{z}) & \delta(S^{z}) & \varepsilon(S^{z}) \\
\delta(S^{z}) & \beta(S^{z}) & \eta(S^{z}) \\
\varepsilon(S^{z}) & \eta(S^{z}) & \gamma(S^{z})
\end{array}
\right)\,.
\label{matrix2}
\end{equation}
with
\begin{eqnarray}
\alpha(S^{z}) & = & \langle S^{z}+1,1|\rho|S^{z}+1,1\rangle\\
\beta(S^{z}) & = & \langle S^{z},0|\rho|S^{z},0\rangle\\
\gamma(S^{z}) & = & \langle S^{z}-1-1|\rho|S^{z}-1,-1\rangle\\
\delta(S^{z}) & = & \langle S^{z},1|\rho|S^{z}+1,0\rangle\\
\varepsilon(S^{z}) & = & \langle S^{z}+1,-1|\rho|S^{z}-1,1\rangle\\
\eta(S^{z}) & = & \langle S^{z},-1|\rho|S^{z}-1,0\rangle
\end{eqnarray}
Unfortunately, the evaluation of the matrix elements in
(\ref{matrix2}) for general values of $S$ and $S^{z}$
turns out to be extremely tedious because 
the form of the Clebsch-Gordan coefficients
is drastically more complicated than in the previous case $S_{2}=1/2$.
For simplicity we therefore concentrate on the case $S=1$ where
the only block of the form (\ref{matrix2}) corresponds to
$S^{z}=0$ with $\alpha=\gamma$ and $\delta=\eta$. In this case we have 
the eigenvalues
\begin{eqnarray}
\mu_{1} & = & \frac{1}{30}+\frac{3}{10}G+\frac{2}{15}H\\
\mu_{2} & = & \frac{1}{6}-\frac{1}{2}G\\
\mu_{3} & = & \frac{1}{3}-\frac{2}{3}H
\end{eqnarray}
The eigenvalue $\mu_{1}$ is always positive and corresponds to the
largest $K$-multiplet. Thus, 
$\rho^{T_{2}}$ has negative eigenvalues if and only if
$G>1/3$ or $H>1/2$. Expressed in terms of correlators these conditions read
\begin{eqnarray}
2 & < & \langle(\vec S_{1}\vec S_{2})^{2}\rangle \\
1 & > & 
\langle\vec S_{1}\vec S_{2}\rangle+\langle(\vec S_{1}\vec S_{2})^{2}\rangle
\end{eqnarray}
Similarly as in the previous case $S_{1}=S$, $S_{2}=1/2$, the two-spin
density matrix has a non-positive partial transpose only if the weight
of the smaller multiplets is sufficiently large. We note that, differently
from the case $S_{1}=S$, $S_{2}=1/2$, an analogous  proof for the 
sufficiency of the Peres-Horodecki criterion cannot be given for 
$S_{2}\geq 1$ since such states are determined by more than just a single
correlator.

\section{Conclusions}

We have analyzed the pairwise quantum entanglement 
in SU(2)-invariant
quantum spin systems by studying their behavior under partial 
transposition. As a general result, the partial transpose 
of an SU(2)-invariant two-spin state has the same 
multiplet structure and degeneracies 
as the original matrix with eigenvalue of largest multiplicity being
non-negative. 
SU(2)-invariant density matrices arise from thermal equilibrium states of 
isotropic spin systems in sufficiently low spatial dimension.
The case $S_{1}=S$, $S_{2}=1/2$ can be
solved completely and is discussed in detail with respect to well-known 
Heisenberg spin models. Moreover, in this case the Peres-Horodecki 
ciriterion turns out to be a sufficient condition for non-separability.
As a general trend, low spatial dimension
and small lengths of the spins involved tend to facilitate the
occurrence of inseparable equilibrium states. However, inseparability
occurs typically only between neighboring spins in such spin lattice systems.
In this sense, isotropic Heisenberg spin models do not appear to be a
particularly strong source of quantum entanglement, at least as far as
their equilibrium properties are concerned. Finally we have also
characterized the properties of SU(2)-invariant states of
two spins of length 1 under partial transposition.

\acknowledgments{
I thank Maciej Lewenstein and R.~F. Werner for useful discussions,
and Guo-Qiang Zhu for pointing out typos in the manuscript.}


\begin{references}
\bibitem{note2}
As it was noticed in \protect{\cite{Vollbrecht01}} there is
a slight ambiguity in the construction of $U_{1}\otimes U_{2}^{*}$
concerning a global phase. This phase, however, does not influence the action
of $U_{1}\otimes U_{2}^{*}$ on any operator.
\end{references}
\end{document}